\documentclass[]
  {aipproc}

\layoutstyle{6x9}

\begin{document}

\title 
      [Infandum, regina, iubes renovare dolorem]
      {Constraining the properties of dark energy}

\iftrue
\author{Dragan Huterer}{
  address={Physics Department, University of Chicago, Chicago, IL},
  email={dhuterer@sealion.uchicago.edu},
}

\author{Michael S. Turner}{
  address={Physics and Astronomy and Astrophysics Departments, 
	University of Chicago, Chicago, IL},
  altaddress={When I go to work}{NAG Ltd, Oxford}
  email={mturner@oddjob.uchicago.edu},
}
\fi

\begin{abstract}
The presence of dark energy in the Universe is inferred directly
from the accelerated expansion of the Universe, and indirectly,
from measurements of cosmic microwave background (CMB)
anisotropy.  Dark energy contributes about 2/3 of the critical
density, is very smoothly distributed, and has large negative
pressure.  Its nature is very much unknown.  Most of its
discernible consequences follow from its effect on evolution of
the expansion rate of the Universe, which in turn affects the
growth of density perturbations and the age of the Universe, and
can be probed by the classical kinematic cosmological tests. Absent a
compelling theoretical model (or even a class of models), we
describe dark energy by an effective equation of state
$w=p_X/\rho_X$ which is allowed to vary with time.   We
describe and compare different approaches for determining $w(t)$,
including magnitude-redshift (Hubble) diagram, number counts of
galaxies and clusters, and CMB anisotropy, focusing particular
attention on the use of a sample of several thousand type Ia
supernova with redshifts $z< 1.7$, as might be gathered by
the proposed SNAP satellite. Among other things, we derive
optimal strategies for constraining cosmological parameters using
type Ia supernovae. While in the near term CMB anisotropy will
provide the first measurements of $w$, supernovae and number
counts appear to have the most potential to probe dark energy. 

\end{abstract}

\date{\today}

\maketitle

\section{Introduction}

Three major lines of evidence point to the existence of a smooth
energy component in the universe. Various measurements of the
matter density indicate $\Omega_M\simeq 0.3\pm 0.1$ (e.g.,
~\cite{mst_scripta}). Recent cosmic microwave background (CMB) results
strongly favor a flat (or nearly flat) universe, with the total
energy density $\Omega_0\simeq 1.1\pm 0.07$~\cite{jaffe}. 
Finally, there is direct evidence coming from type Ia
supernovae (SNe Ia) that the universe is accelerating its
expansion, and that it is dominated by a component with strongly
negative pressure, $w=p_X/\rho_X< -0.6$~\cite{riess,
perlmutter-1999}. Two out of of three of these arguments would
have to prove wrong in order to do away with the smooth
component.
 
Even before the direct evidence from SNe Ia, there was a dark
energy candidate: the energy density of the quantum vacuum (or
cosmological constant) for which $p=-\rho$.  However, the
inability of particle theorists to compute the energy of the
quantum vacuum -- contributions from well understood physics
amount to $10^{55}$ times critical density -- casts a dark shadow
on the cosmological constant. Another important issue is the
coincidence problem: dark energy seems to start dominating
the energy budget, and accelerating the expansion of the
universe, just around the present time.  A number of other
candidates have been proposed: rolling scalar field (or
quintessence), and network of frustrated topological defects, to
name a couple.  While these and other models have some motivation
and attractive features, none are compelling.

In this work we discuss the cosmological consequences of dark
energy that allow its nature to be probed. We also discuss
relative merits of various cosmological probes, focusing
particular attention to a large, well-calibrated sample of SNe Ia
that would be obtained by the proposed space telescope
SNAP\footnote{http://snap.lbl.gov}. We parameterize dark
energy by its scaled energy density $\Omega_X$ and
equation of state $w$, and assume a fiducial cosmological
model with $\Omega_X=1-\Omega_M=0.7$ and $w=-1$, unless indicated
otherwise.

\section{Cosmological consequences of the Dark Energy}

Dark energy is smooth, and if does clump at all, it does so at very
large scales only ($k\sim H_0$). All of its consequences therefore
follow from its modification of the expansion rate

$$
H(z)^2   =  H_0^2\left[ \Omega_M(1+z)^3 +
\Omega_X \exp [3\int_0^z\,(1+w(x))d\ln (1+x)] \right].
$$

For a given rate of expansion today $H_0$, the expansion rate
in the past, $H(z)$, was smaller in the presence of dark
energy. Therefore, dark energy increases the age of the
universe. The comoving distance $r(z)=\int dz/H(z)$ also
increases in the presence of dark energy. The same follows
for the comoving volume element $dV/d\Omega\,dz=r^2(z)/H(z)$.

\begin{figure}
  \includegraphics[height=.3\textheight]{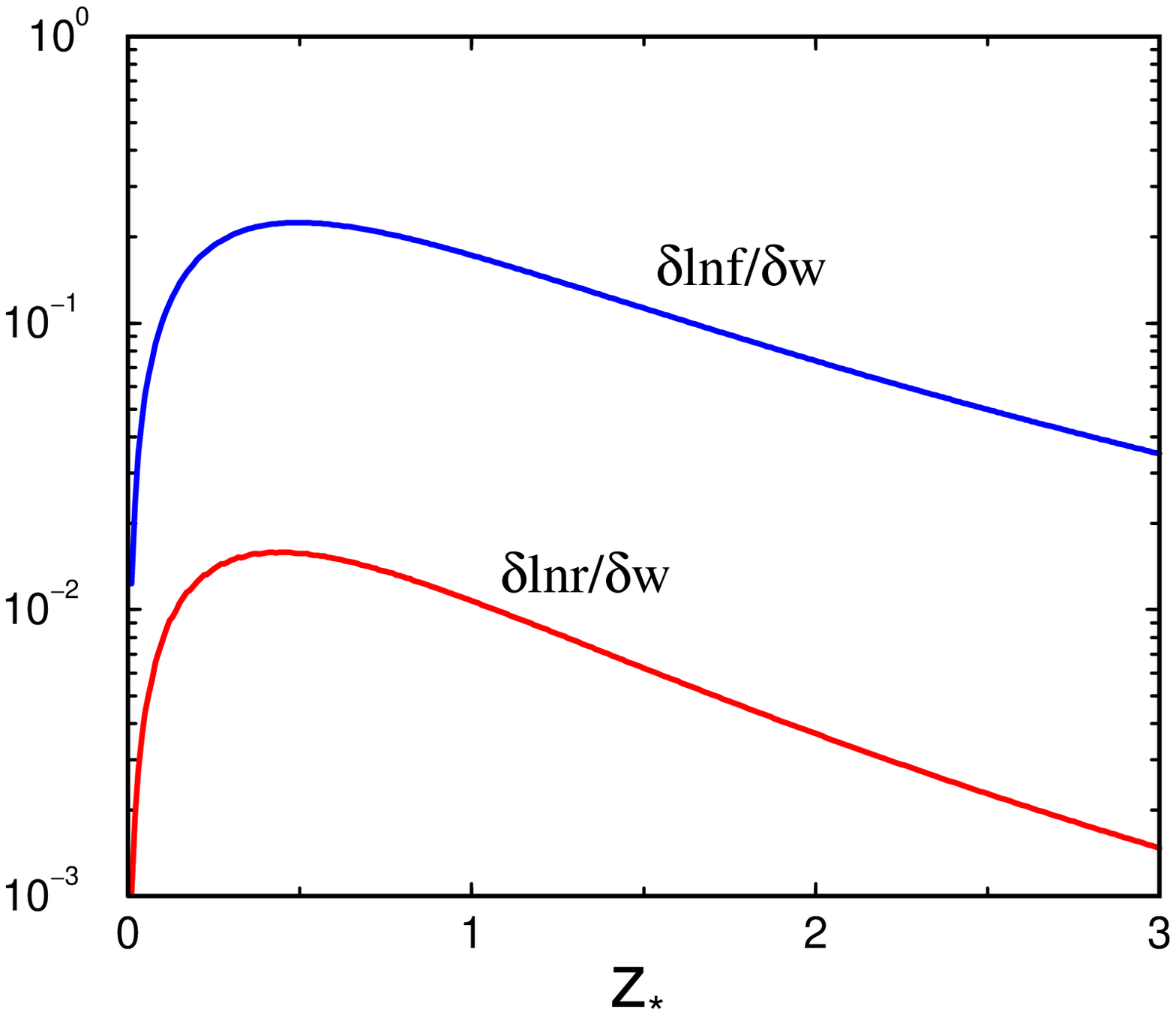}
  \includegraphics[height=.3\textheight]{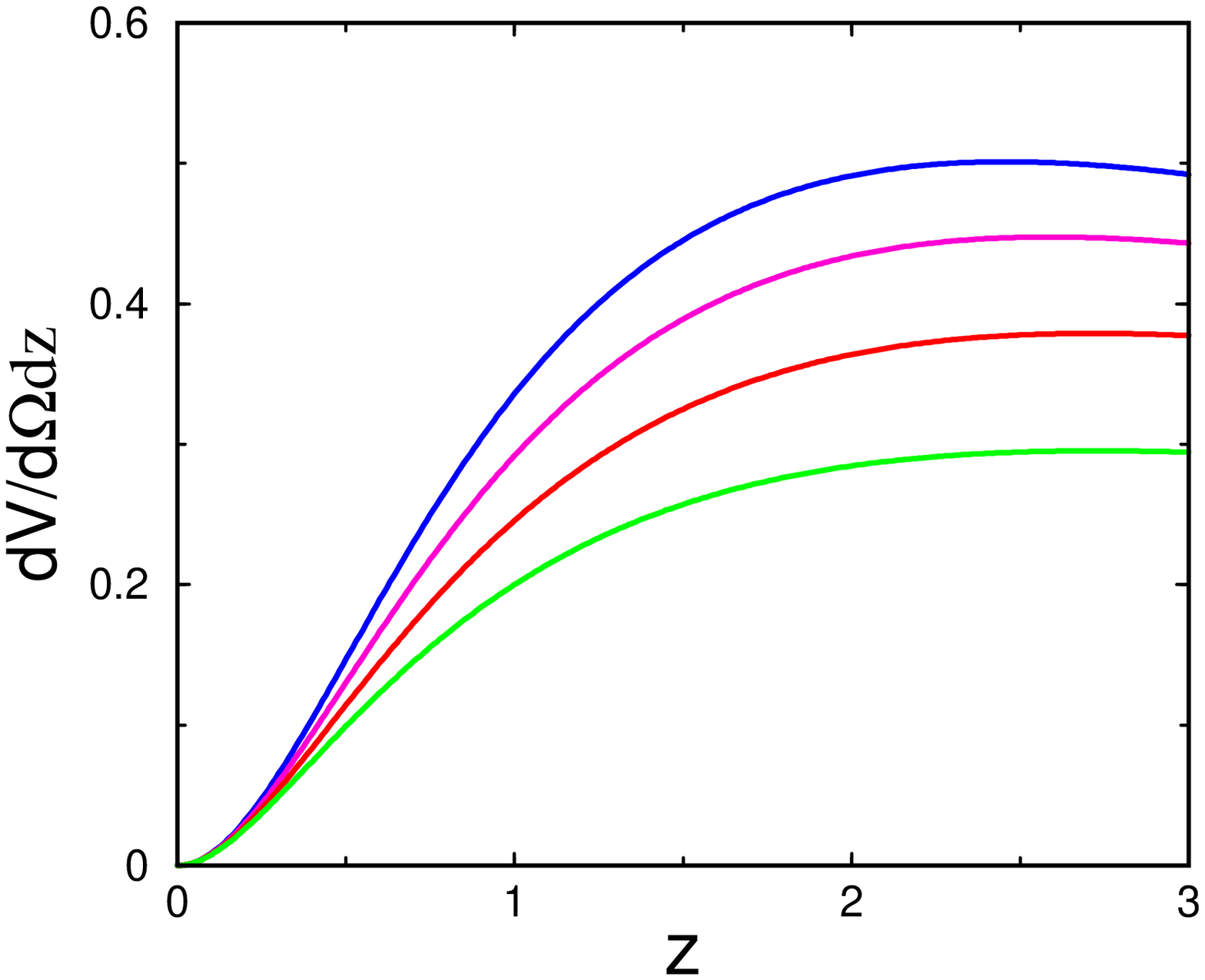}

  \caption{Left panel: relative sensitivity of the comoving
distance $r(z)$ and the comoving volume element $f(z)\equiv
dV/d\Omega\,dz$ to a localized change in the value of $w$ at
redshift $z_*$. Right panel: comoving volume element $dV/d\Omega\,
dz$ vs.\ redshift for constant $w=-1, -0.8, -0.6, -0.4$ (from top
to bottom).}
\label{fig:kinematic}
\end{figure}

At small redshifts $r(z)$ is insensitive to $w$ for
the simple reason that {\em all} cosmological models
reduce to the Hubble law ($r = H_0^{-1} z$) for $z\ll 1$:

\begin{equation}
r(z) \rightarrow H_0^{-1}\left[ z -{3\over 4}z^2 -{3\over 4}
\Omega_X wz^2 +\cdots \right] \ \ {\rm for}\ z\ll 1.
\label{eq:hubblelaw}
\end{equation}
At redshift greater than about five, the sensitivity of $r(z)$ to
a change in $w$ levels off because dark energy becomes an
increasingly smaller fraction of the total energy density,
$\rho_X /\rho_M \propto (1+z)^{3w}$.  Note also that the volume
depends upon dark energy parameters through $H(z)$, which
contains one integral less than the distance $r(z)$. This is the
reason why number-count surveys (which effectively measure
$dV/d\Omega\, dz$) are potentially a strong probe of dark
energy.
 
\subsection{Constraints on (constant) w}

Supernovae are perhaps the strongest probes of dark energy,
as the relevant observable $r(z)$ depends only on $\Omega_M$,
$\Omega_X$ and $w$; moreover, SNe probe the optimal redshift
range.  A powerful supernova program, such as SNAP, with about
2000 SNe distributed at $0.2<z<1.7$, would be able to determine
$w$ with an accuracy $\sigma_w=0.05$.

\begin{figure}
  \includegraphics[height=.3\textheight]{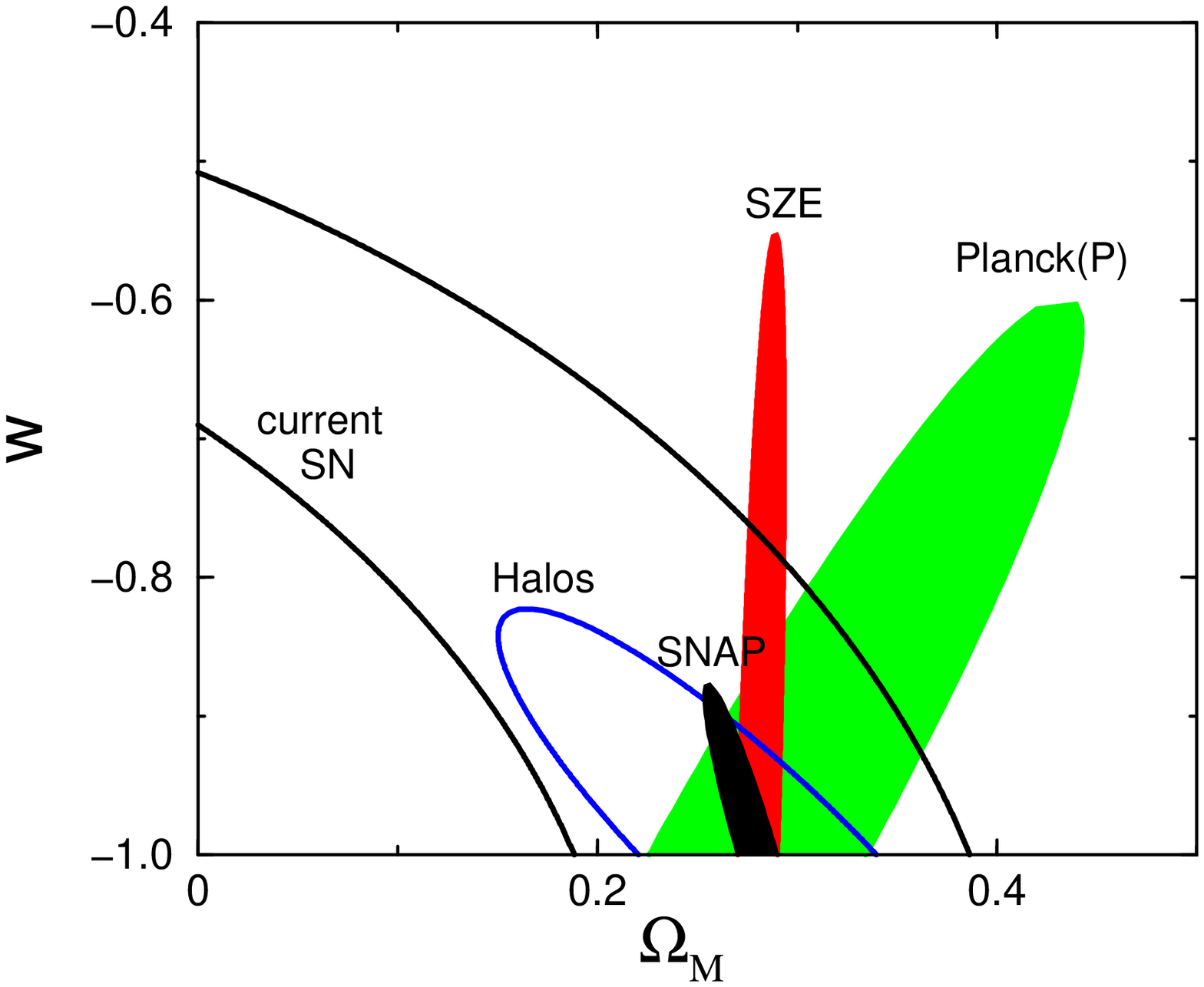}
  \includegraphics[height=.3\textheight]{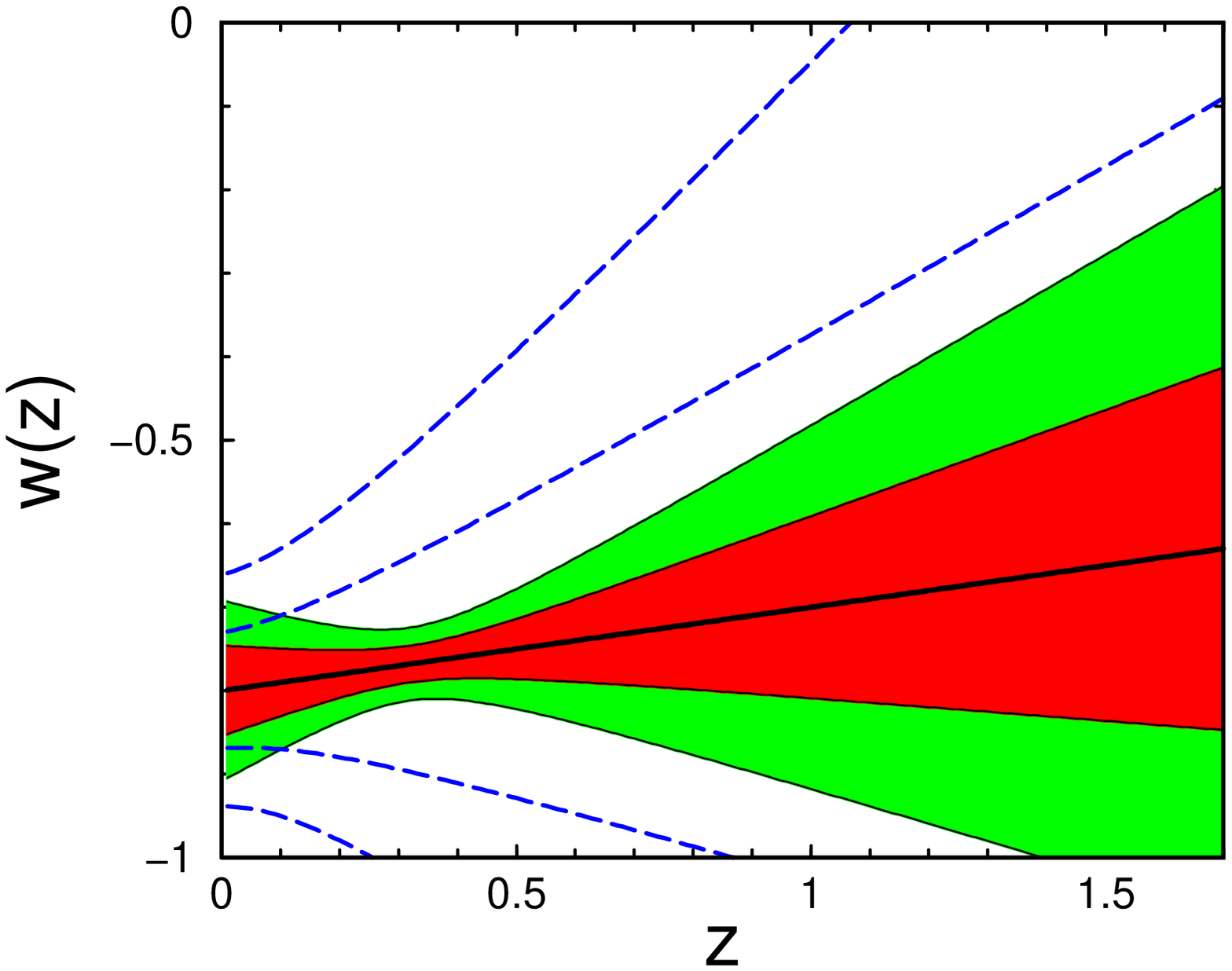}
  \caption{Left panel: present and projected future constraints
  (68\% C.L.) on constant $w$. The Sunyaev-Zeldovich effect (SZE)
  constraint roughly corresponds to the estimate
  from~\cite{holder}, while for the halo counts we assumed a
  random uncertainty of 10\% in each of eight redshift
  bins. Right panel: projected constraints (68\% and 95\% C.L.)
  on time-varying $w$ using simulated data expected from
  SNAP. The fiducial model is $w(z)=-0.8+0.1\,z$ and $\Omega_M$
  is considered to be known. The broken lines show the effect of
  assuming a Gaussian uncertainty of $0.05$ in $\Omega_M$.}
  \label{fig:CMB}
\end{figure}

Number-count surveys are potentially also a very strong probe of
dark energy. To estimate the expected number density of objects
(galaxies, galaxy clusters etc.), one typically uses the
Press-Schechter formalism~\cite{PS}.  For example, number counts of galaxy
clusters either from an X-ray survey or from Sunyaev-Zeldovich
survey can serve as a probe of dark energy~\cite{holder}. The
number of these objects depends upon the growth of structure as
well as on cosmological volume, and the former makes the overall
constraint complementary to that of SNe Ia. Another example is
using halos of fixed rotational speed~\cite{davis} to infer the
dark energy dependence of the volume element. In order to obtain
constraints on dark energy using number-count surveys,
control over modeling and systematic errors will be critical.

CMB anisotropy, which mainly probes the epoch of recombination
($z\sim 1000$) is weakly sensitive to the presence of dark
energy because dark energy becomes inconsequential at such
high redshift. The small dependence comes through the distance to
the surface of last scattering  which slightly
increases in the presence of dark energy, moving the acoustic
peaks to smaller scales. However, this effect is small:

$$
{\Delta l_1\over l_1} = -0.084 \Delta w -0.23{\Delta \Omega_Mh^2
	\over \Omega_Mh^2}
        +0.09{\Delta \Omega_Bh^2\over \Omega_Bh^2} 
        +0.089{\Delta \Omega_M \over \Omega_M}
        -1.25{\Delta \Omega_0 \over \Omega_0},
\label{eq:cmb_firstpeak}
$$

\noindent which shows that the location of the first peak is
least sensitive to $w$.  The left panel of Fig.~\ref{fig:CMB}
further illustrates this: even the Planck experiment with
polarization information would be able to achieve only
$\sigma_w\approx 0.25$ (after marginalization over all other
parameters).  Nevertheless, CMB constraints are of crucial
importance because they complement SNe and other probes, in
particular providing the measurement of the total energy density
$\Omega_0$.

The Alcock-Paczynski shape test and the age of the Universe are
also sensitive to the presence of dark energy, but they seem
somewhat less promising: the former because of the small size of
the effect (around 5\%); and the latter because the errors in the
two needed quantities, $H_0$ and $t_0$, are not likely to become
small enough in the near future. Finally, large-scale structure
surveys have weak direct dependence upon dark energy simply
because this component is smooth on observable scales.

\subsection{Probing $w(t)$}

There is no {\it a priori} reason to assume constant $w$, as some
models (in particular quintessence) generically produce
time-varying $w$. Here we discuss how to best constrain $w(t)$
(or $w(z)$). We find that $w(z)$ will be much more difficult to
constrain than constant $w$ due to additional degeneracies. In
order to illustrate prospects for constraining $w(z)$, we use
SNAP's projected dataset with 2000 SNe and assume that, by the
time this difficult task is seriously attempted, $\Omega_M$ and
$\Omega_X$ will be pinned down accurately by a combination of CMB
measurements and large-scale structure surveys.

One of the simplest ways to characterize $w(z)$ is to divide the
redshift range into $B$ redshift bins and assume constant
equation of state ratio $w_i$ in each. The resulting constraint,
however, is poor for $B\geq 3$ (and ideally one would want many
more bins). A better way to proceed is to assume linearly varying
$w(z)$ expanded around a suitably chosen redshift
$w_1$~\cite{coorayhuterer}

$$
w(z)=w_1 +w'(z-z_1).
$$

We choose $z_1$ so as to de-correlate $w_1$ and $w'$. The
resulting constraint is shown in the right panel of Fig.
~\ref{fig:CMB}. The constraint is best at $z\approx 0.4$ and
deteriorates at lower and higher redshifts.  Despite the
relatively large uncertainty in the slope ($\sigma_{w'}=0.16$),
this analysis may be useful in constraining dark energy
models.  The parameterization $w(z)=w_1 -\alpha
\ln[(1+z)/(1+z_1)]$ yields comparable constraints. 

\section{Optimal Strategies}

Given the importance and difficulty of probing dark energy, it
is worthwhile to consider how to optimize data sets in order to
obtain tighter constraints.  We address the following
question: what is the optimal redshfit distribution of SNe Ia in
order to best constrain the cosmological parameters? 

\begin{figure}
  \includegraphics[height=.3\textheight]{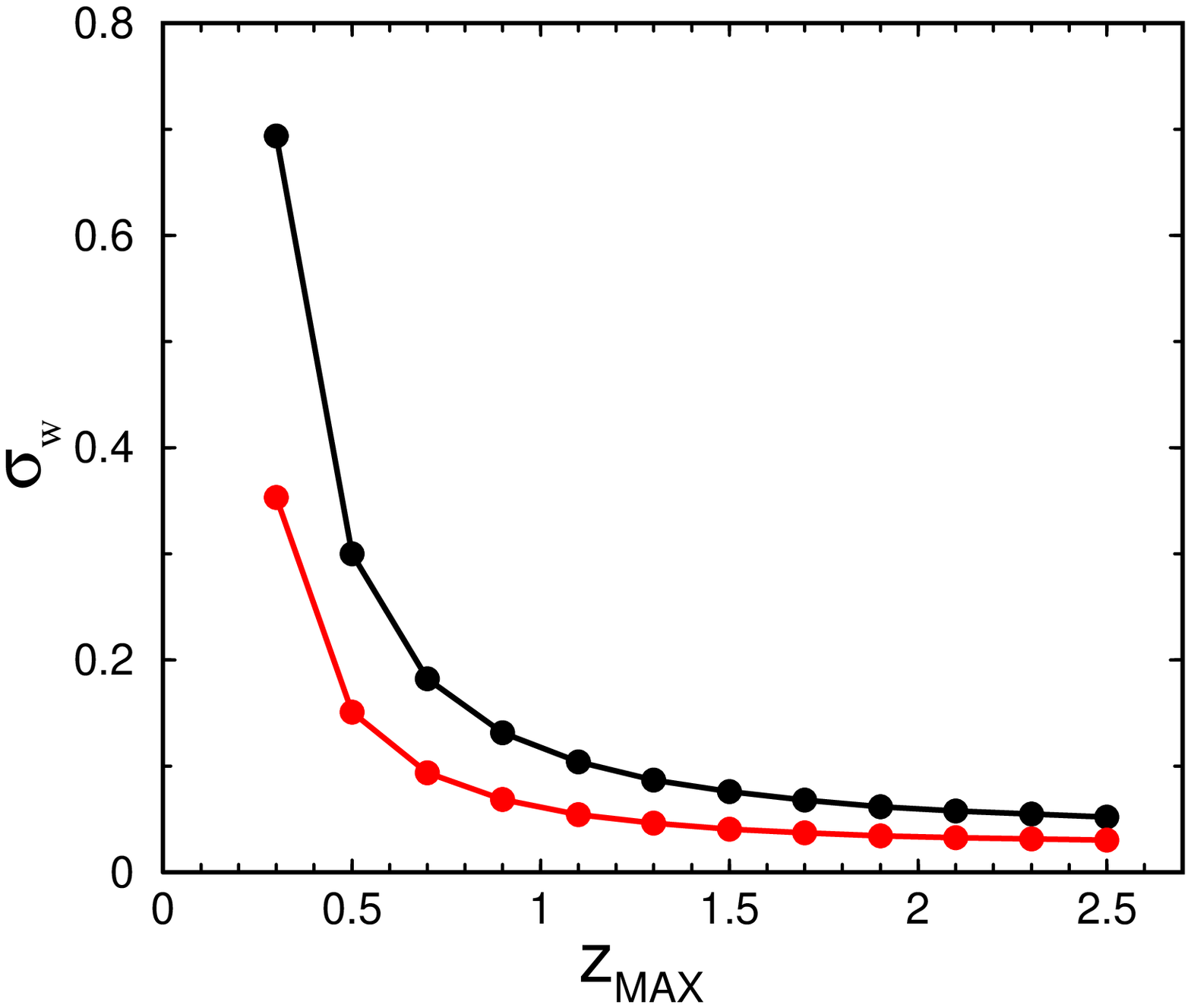}
  \includegraphics[height=.3\textheight]{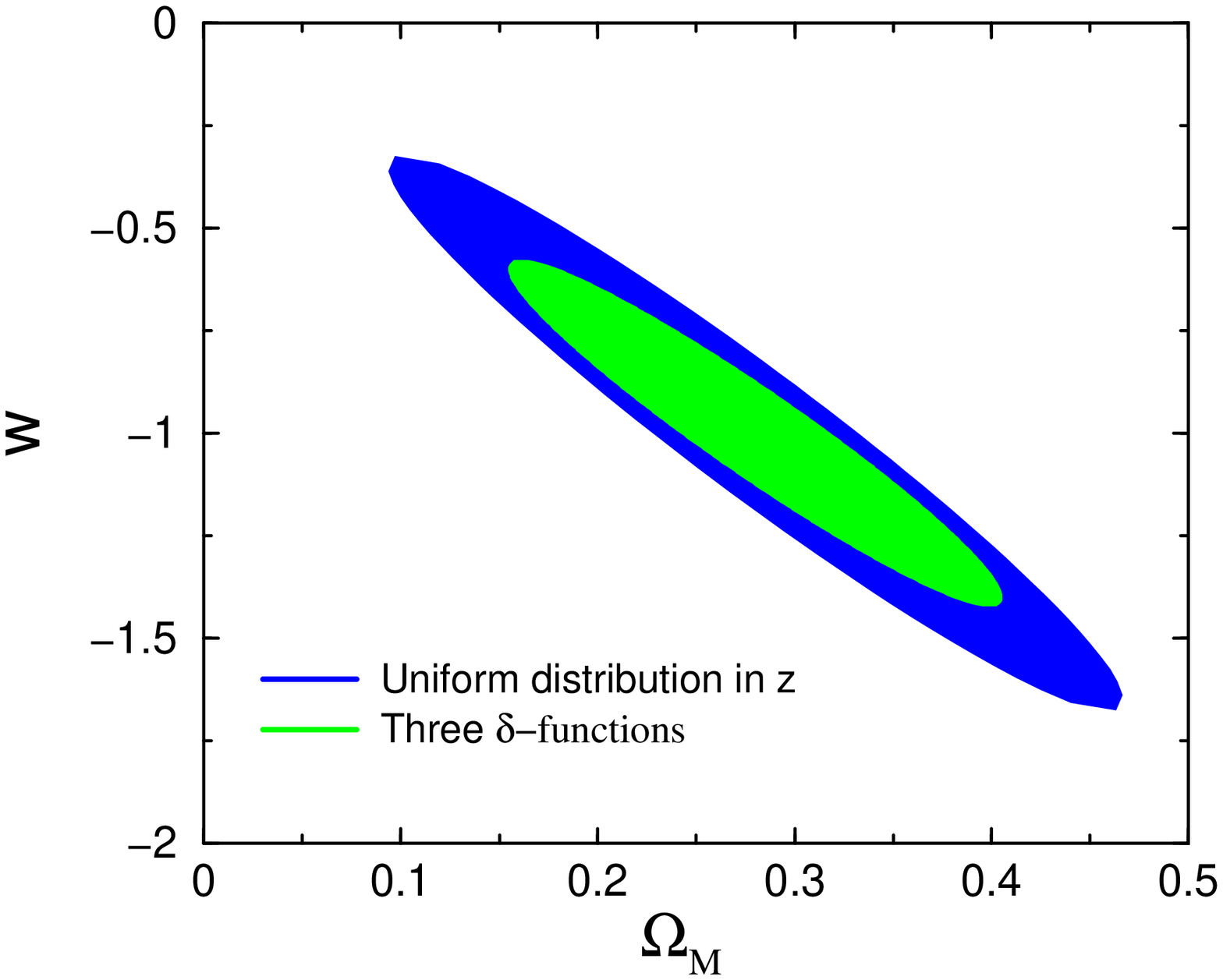}
  \caption{Left panel: upper curve shows the constraint upon
  (constant) $w$ using the (nearly uniform) distribution of 2000
  SNe out to $z_{\rm max}$. Lower curve shows uncertainties using
  the same number of SNe with mathematically optimal distribution
  in redshift. Right panel: constraints on $\Omega_M$ and $w$
  using 100 SNe with uniform (dark) and optimal (light)
  distribution in redshift. } \label{fig:optimal}
\end{figure}

We choose to minimize the area of uncertainty ellipsoid for $P$
parameters, which (in the Fisher matrix
formalism~\cite{Fisher-Tegmark}) is equivalent to maximizing
$\det F$, where $F$ is the Fisher matrix.  Making a few
simplifying but reasonable assumptions, in~\cite{sn_optimal} we
show that $P$ parameters are always best determined if SNe are
distributed as $P+1$ delta-functions in redshifts, two of which
are always at the lowest and highest redshifts available. For
example, to best constrain $\Omega_X$ and $w$ (assuming a flat
universe), SNe should be located at $z=0$, $z\approx 2/5 \,z_{\rm
max}$ and $z=z_{\rm max}$ in equal numbers, where $z_{\rm max}$
is the maximum redshift available. In Fig.~\ref{fig:optimal} we
show the merits of optimal distribution. Note, however, that
there are other considerations that may favor a more uniform
distribution of SNe in redshift -- for example, ability to
perform checks for dust and evolution.

\section{Conclusions}

We discussed prospects for constraining dark energy.  Dark
energy is best probed at redshifts roughly between $0.2$ and
$1.5$, where it is dynamically important. Probes of the
low-redshift Universe (supernovae and number counts) seem the
most promising, as they only depend upon three cosmological
parameters ($\Omega_M$, $\Omega_X$ and $w$), which will be
effectively reduced to two ($\Omega_X$ and $w$) when precision
CMB measurements determine $\Omega_0 =\Omega_M +
\Omega_X$ to better than 1\%. CMB is weakly sensitive to the 
presence of dark energy, but is very important as a complementary
probe. Alcock-Paczynski test and the age of the universe seem less
promising.

Constraining the equation of state $w$ is the first step in
revealing the nature of the dark energy. We show that future
surveys will be able to determine constant $w$ quite
accurately. A high-quality sample of 2000 supernovae out to
redshift $z\sim 1.7$ could determine $w$ to a precision of
$\sigma_w =0.05$. A similar accuracy might be achieved by number
counts of galaxies and clusters of galaxies out to $z\sim 1.5$,
provided systematics are held in check (e.g., in the case of
galaxies, the evolution of the comoving number density needs to
be known to better than 5\%).  Time-varying $w$ will be
considerably more difficult to pin down because of additional
degeneracies that arise in this case.  Nevertheless, interesting
constraints may be achieved using high-quality data and assuming
a slowly-varying $w(z)$.  Consequently, use of complementary
measurements and further development of techniques to constrain
dark energy will be of crucial importance.
 
\begin{theacknowledgments}
We would like to thank the organizers for the wonderful
conference and the ``black hole'' drink recipe. 
This work was supported by DOE at Chicago. 
\end{theacknowledgments}


\bibliography{proceed}

\end{document}